\documentclass[11pt,a4paper]{article}
\usepackage[latin1]{inputenc}

\usepackage[english]{babel}

\usepackage{amsmath}
\usepackage{amsfonts}
\usepackage{amssymb}
\usepackage{makeidx}
\usepackage{graphicx}
\usepackage{xcolor}
\usepackage[T1]{fontenc}
\usepackage{enumitem}

\usepackage[round]{natbib}
\usepackage[colorlinks]{hyperref}
\usepackage{hyperref}
\hypersetup{colorlinks,
            citecolor=blue
}
\usepackage[left=2cm,right=2cm,top=2cm,bottom=2cm]{geometry}

\title{Stability of Planetary Motion in Binary Star Systems
}

\author{R. Capuzzo-Dolcetta, Dip. di Fisica, Sapienza, Univ. of Roma, Italy; \\ G. De Cesare, INAF-OAS, Bologna, Italy; A. Marino, INAF-OAR, Monteporzio Catone, Italy}
\date{}
\begin{document}
\maketitle
%\author{Sara Rastello}
%\date{\vspace{-5ex}}

\def\mnras{MNRAS}%
% Bibliography and bibfile
\def\aj{AJ}%
          % Astronomical Journal
\def\araa{ARA\&A}%
          % Annual Review of Astron and Astrophys
\def\apj{ApJ}%
          % Astrophysical Journal
\def\apjl{ApJL}%
          % Astrophysical Journal, Letters
\def\apjs{ApJS}%
          % Astrophysical Journal, Supplement
\def\ao{Appl.~Opt.}%
          % Applied Opt
\def\apss{Ap\&SS}%
          % Astrophysics and Space Science
\def\aap{A\&A}%
          % Astronomy and Astrophysics
\def\aapr{A\&A~Rev.}%
          % Astronomy and Astrophysics Reviews
\def\aaps{A\&AS}%
          % Astronomy and Astrophysics, Supplement
\def\azh{AZh}%
          % Astronomicheskii Zhurnal
\def\baas{BAAS}%
          % Bulletin of the AAS
\def\bac{Bull. astr. Inst. Czechosl.}%
          % Bulletin of the Astronomical Institutes of Czechoslovakia
\def\caa{Chinese Astron. Astrophys.}%
          % Chinese Astronomy and Astrophysics
\def\cjaa{Chinese J. Astron. Astrophys.}%
          % Chinese Journal of Astronomy and Astrophysics
\def\icarus{Icarus}%
          % Icarus
\def\jcap{J. Cosmology Astropart. Phys.}%
          % Journal of Cosmology and Astroparticle Physics
\def\jrasc{JRASC}%
          % Journal of the RAS of Canada
\def\mnras{MNRAS}%
          % Monthly Notices of the RAS
\def\memras{MmRAS}%
          % Memoirs of the RAS
\def\na{New A}%
          % New Astronomy
\def\nar{New A Rev.}%
          % New Astronomy Review
\def\pasa{PASA}%
          % Publications of the Astron. Soc. of Australia
\def\pra{Phys.~Rev.~A}%
          % Physical Review A: General Physics
\def\prb{Phys.~Rev.~B}%
          % Physical Review B: Solid State
\def\prc{Phys.~Rev.~C}%
          % Physical Review C
\def\prd{Phys.~Rev.~D}%
          % Physical Review D
\def\pre{Phys.~Rev.~E}%
          % Physical Review E
\def\prl{Phys.~Rev.~Lett.}%
          % Physical Review Letters
\def\pasp{PASP}%
          % Publications of the ASP
\def\pasj{PASJ}%
          % Publications of the ASJ
\def\qjras{QJRAS}%
          % Quarterly Journal of the RAS
\def\rmxaa{Rev. Mexicana Astron. Astrofis.}%
          % Revista Mexicana de Astronomia y Astrofisica
\def\skytel{S\&T}%
          % Sky and Telescope
\def\solphys{Sol.~Phys.}%
          % Solar Physics
\def\sovast{Soviet~Ast.}%
          % Soviet Astronomy
\def\ssr{Space~Sci.~Rev.}%
          % Space Science Reviews
\def\zap{ZAp}%
          % Zeitschrift fuer Astrophysik
\def\nat{Nature}%
          % Nature
\def\iaucirc{IAU~Circ.}%
          % IAU Cirulars
\def\aplett{Astrophys.~Lett.}%
          % Astrophysics Letters
\def\apspr{Astrophys.~Space~Phys.~Res.}%
          % Astrophysics Space Physics Research
\def\bain{Bull.~Astron.~Inst.~Netherlands}%
          % Bulletin Astronomical Institute of the Netherlands
\def\fcp{Fund.~Cosmic~Phys.}%
          % Fundamental Cosmic Physics
\def\gca{Geochim.~Cosmochim.~Acta}%
          % Geochimica Cosmochimica Acta
\def\grl{Geophys.~Res.~Lett.}%
          % Geophysics Research Letters
\def\jcp{J.~Chem.~Phys.}%
          % Journal of Chemical Physics
\def\jgr{J.~Geophys.~Res.}%
          % Journal of Geophysics Research
\def\jqsrt{J.~Quant.~Spec.~Radiat.~Transf.}%
          % Journal of Quantitiative Spectroscopy and Radiative Trasfer
\def\memsai{Mem.~Soc.~Astron.~Italiana}%
          % Mem. Societa Astronomica Italiana
\def\nphysa{Nucl.~Phys.~A}%          % Nuclear Physics A
\def\physrep{Phys.~Rep.}%
          % Physics Reports
\def\physscr{Phys.~Scr}%
          % Physica Scripta
\def\planss{Planet.~Space~Sci.}%
          % Planetary Space Science
\def\procspie{Proc.~SPIE}%

%\tableofcontents %GENERA INDICE
%\section*{Job Purpose}
%\section*{Job Purpose}

\stepcounter{section}
\begin{abstract}
Many exoplanets have been discovered in binary star systems, on internal (S-type) or circumbinary (P-type) orbits.
In this study, we considered the problem of stability for planets of finite (non negligible) mass. We approached the problem by selecting a huge set of initial conditions for planetary orbits of the S-type, to perform high precision and very extended in time integrations at varying: the planet mass, its distance from the primary star (whose mass is $m_A$), the mass ratio $\mu =m_B/(m_A + m_B)$ for the two stars in the system, and the eccentricity of the initial star motion.
For our numerical integrations, we resorted to the use of a 15th order integration scheme (IAS15, available within the REBOUND framework), that provides an optimal solution for long-term time integrations. We estimated the probability of different types of instability: planet collisions with the primary or secondary star or planet ejected away from the binary star system. We confirm and generalize to massive planets the dependence of the critical semi-major axis on eccentricity and mass ratio of the binary already found by \cite{howi99}.
We were also able to pick a significant number of orbits that are only `marginally' stable, according to the classification introduced by \cite{mu05}.

A, natural, extension of this work has been the study of the effect of perturbations induced to circumbinary planet motion by a passing-by star, like it often happens in a star cluster.
One of the targets of this analysis is the investigation of the possibility that a planet, formerly on a stable S-type orbit around one of the two stars, could transit to a stable P-type orbit (or viceversa). We performed a series of more than 4500 scattering experiments with different initial conditions typical of encounters in small star clusters. We found some interesting behaviors of the systems after perturbation and showed how a transition from an inner (S-type) stable orbit to a circumbinary (P-type) (and vice-versa) has a very low (but non null) probability.
    
\end{abstract}

\section{Introduction}
\begin{figure*}
	%\includegraphics[width=\columnwidth]{fig/planetary_mass}
	\begin{tabular}{cc}
	\includegraphics[width=\columnwidth]{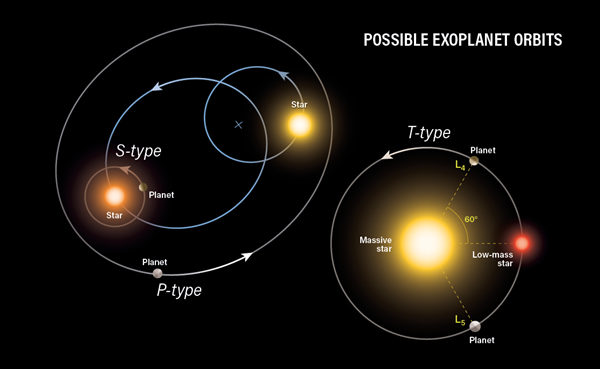}
    \end{tabular}
    \caption{An artistic view of the planetary orbits in a binary star system. Credit: astronomy.com (Roen Kelly).}
    \label{fig:1}
\end{figure*}

Orbits of planets in binary systems (Fig. \ref{fig:1}) are traditionally classified into three categories \citep{dvo86}: 

\begin{enumerate}[label=(\roman*)]
\item the Planet-type (P-type) external orbits around both stars in the binary,
\item the Satellite type (S-type) internal orbits, around one of the two stars, and 
\item the Libration-type (L-type) orbits, corresponding to librations around the  Lagrangian equilibrium points L$_4$ and L$_5$, which are stable when the stellar mass ratio $m_B/(m_A+m_B)$ is less than $\sim 0.04$ (assuming $m_B < m_A$). 
\end{enumerate}

While L-type orbits are not of practical interest for exoplanets in binary systems, P-type and S-type orbits are indeed of relevant astronomical impact and deserve to be deeply studied in their characteristics.
In this note we report some outcomes of the numerical study of: 
\begin{itemize}
    \item 
the stability of S-type orbits of planets with a negligible mass as well as with a mass of the order of one or 30 Jupiter mass;
\item
the consequences of a star scattering with the planetary system, with different initial conditions typical of encounters in small star clusters.
\end{itemize}
Two publications with more details and results on these topics are in preparation \citep{decd20,madecd20}.

\section{Model and Methods}

We approach the problem of the planet orbital stability in a numerical way, by integrating the orbits under different initial conditions. We consider a system made up by two star $A$ (called primary) and $B$ (called secondary) in motion around their center of mass, with the planet in a co-planar orbit with zero initial eccentricity.  We call \emph{unperturbed} orbit the orbit around the primary star that the planet would have without the presence of the secondary star; in our configuration the  unperturbed orbit is circular. 
We varied: i) the mass ratio of the 2 stars, ii) their orbital eccentricity; iii) the initial planetary orbit radius; iv) the planet mass.
\begin{figure*}
	%\includegraphics[width=\columnwidth]{fig/planetary_mass}
	\begin{tabular}{cc}
	\includegraphics[width=\columnwidth]{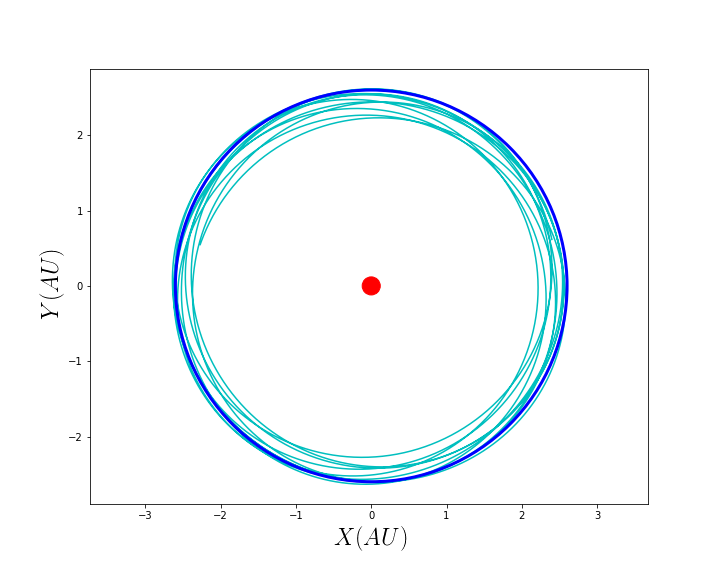}
    \end{tabular}
    \caption{An example stable planetary orbity (cyan) shown in the frame of the primary star (red circle) . The secondary star, at the distance of 10 AU from the primary star, is not shown. The darker blue circle is the unperturbed orbit, i.e. the orbit of the planet if there was no secondary star.}
    \label{fig:2}
\end{figure*}
\begin{figure*}
	%\includegraphics[width=\columnwidth]{fig/planetary_mass}
	\begin{tabular}{cc}
    \includegraphics[width=\columnwidth]{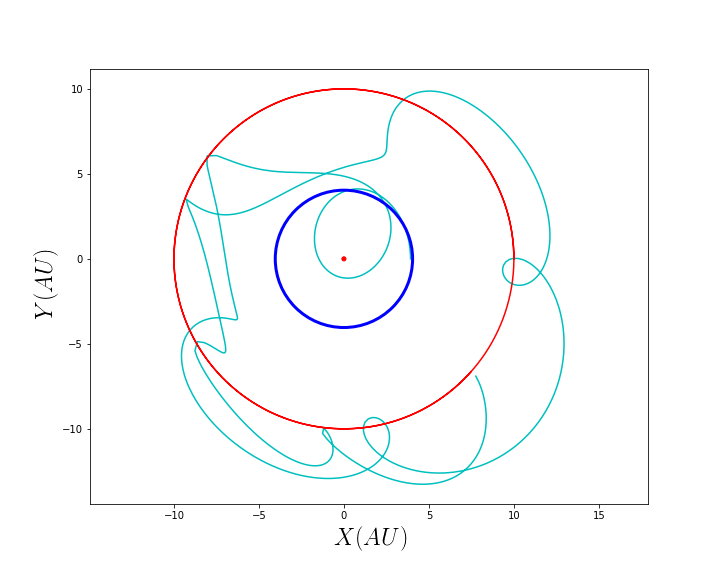}
    \end{tabular}
    \caption{An example of collision of the planet (cyan) with the secondary star (red).}
    \label{fig:3}
\end{figure*}
By putting all these degrees of freedom together, our dataset is made by more than ten thousand orbits.
To obtain, at the same time, an acceptable calculation time and accuracy we use the  high-precision (15th order) integration scheme \emph{IAS15} \citep{Rein2015}. With \emph{IAS15} the systematics errors are kept below machine precision for long term integration over at least $10^9$ orbital time scales. \emph{IAS15} is an integration options available in REBOUND \cite{Rein2012}, a modern open source tool for gravitational dynamics. REBOUND is written in standard C99; an easy to use and convenient Python wrapper is also provided.
Unstable orbits are defined as those that lead the planet either to a \textit{collision} onto one of the two stars or to \textit{escape} from the system. To pick unstable orbits we adopted an efficient practical quantitative criterion based on the variation of the planet trajectory respect to the initial one. 

 As a demonstration, we also created a prototype of a real time orbit simulator 
\par\noindent (link to https://giovixo.github.io/exoplanets/index.html).

\begin{figure*}
	%\includegraphics[width=\columnwidth]{fig/planetary_mass}
	\begin{tabular}{cc}
    \includegraphics[width=\columnwidth]{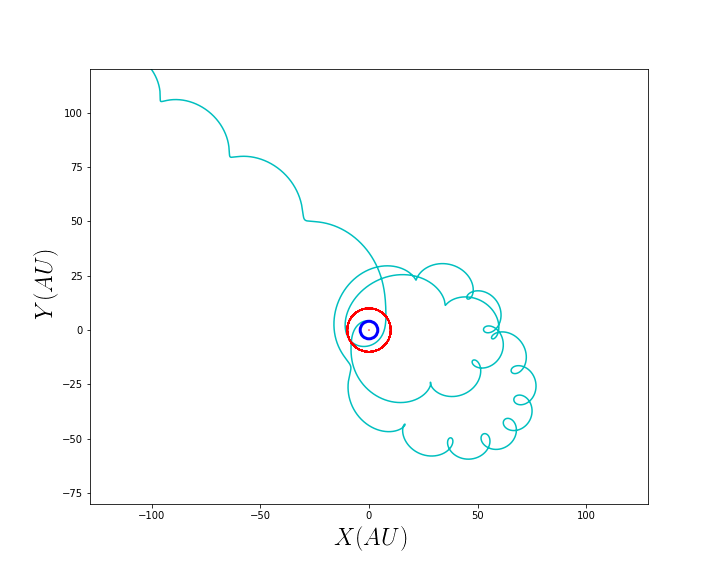}
    \end{tabular}
    \caption{An example of ejection of the planet (cyan) away from the binary.}
    \label{fig:4}
\end{figure*}
\begin{figure*}
	\begin{tabular}{c}
	\includegraphics[width=\columnwidth]{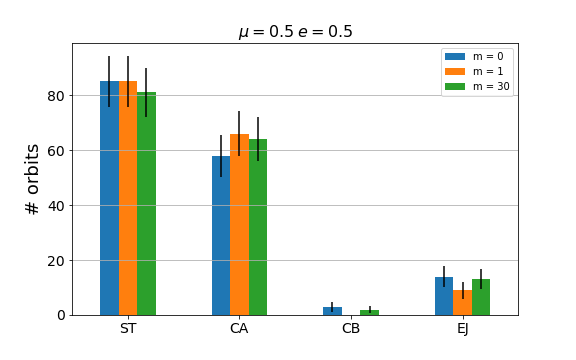}
	\end{tabular}
    \caption{Synopsis of the stability for three different values of the planet mass (0, 1, 30 Jupiter mass) in a binary with $\mu = 0.5, e = 0.5$. The labels are: ST for stable and marginally stable orbits, CA and CB for collisions with the primary or secondary star, EJ for ejections.}
    \label{fig:5}
\end{figure*}
\begin{figure*}
	\begin{tabular}{c}
	\includegraphics[width=\columnwidth]{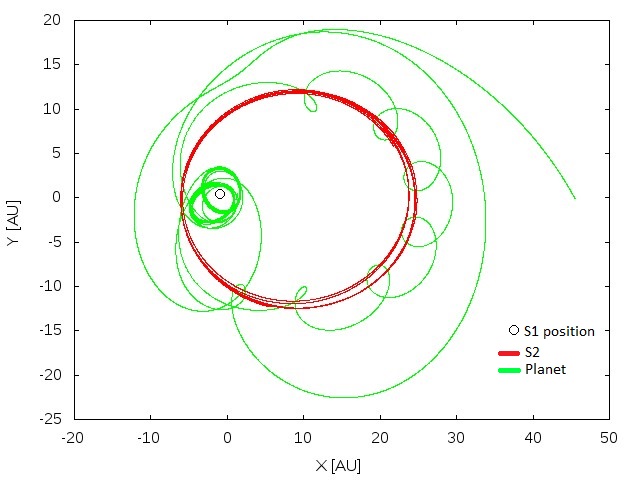}
	\end{tabular}
    \caption{A planetary system perturbed by a passing by star. S1 and S2 are the primary and secondary stars, planet trajectory is in green color. The simulation reported goes from 400 yrs to 1070 yrs since the beginning of the simulation.}
    \label{fig:6}
\end{figure*}

\section{Results}
\label{sec:results}

\label{sec:results}

    We now report how our simulated planetary orbits depend on the initial conditions for the mass and initial semi-major axis of the planet, and on the mass ratio and eccentricity of the binary star system.

 We classify the planetary orbits on the basis of their stability, considering an orbit unstable if (i) the planet will collide in a finite time with one of the two stars or (ii) it is ejected from the binary system. For all the unstable orbits, we also get the \textit{survival} time, i.e. the time elapsed from the beginning of the orbit to the eventual collision or escape. For stable orbits, we estimate the mean deviation $\Delta r$ of the actual orbit from the one unperturbed by the secondary star.
An example of stable planetary orbit is shown in Fig. \ref{fig:2}.\\
Figure \ref{fig:3} shows an orbit leading to a planet collision with the secondary star, while Fig. \ref{fig:4} shows a planet ejection away from the binary. In the last case, we notice that, before escaping, the planet moves in an outer orbit for a while. This leads to the hypothesis that, under certain conditions, a transition from an internal (S-type) orbit to an external (P-type) orbit could actually occur.

The distribution of collisions, ejections and stable orbits is reported in Fig. \ref{fig:5}, for a binary star with $e = 0.5$ and $\mu = 0.5$. In this particular configuration, we find that the statistics of collisions and ejections do not depend on the planetary mass.

\begin{table*}
    \centering
    \begin{tabular}{|c c c c c c c c c c c|}
    \multicolumn{11}{c}{Critical semi-major axis ($m_p = 0, 30$ M$_\textrm{J}$)} \\
    \hline 
    \multicolumn{6}{|c}{} & \multicolumn{1}{c}{$\mu$} & \multicolumn{4}{c|}{} \\
            &    & 0.10        & 0.20       & 0.30        & 0.40         & 0.50         & 0.60       & 0.70       & 0.80        & 0.90 \\
    $e$ & $m$  (M$_\textrm{J}$) & \multicolumn{8}{c}{} & \multicolumn{1}{c|}{} \\   
    \cline{3-11}      
    0.0     & 30 & 0.43        & 0.36       & 0.38        & 0.31         & 0.27         & 0.23       & 0.20       & 0.17        & 0.13 \\
            &  0 & 0.45        & 0.38       & 0.37        & 0.30         & 0.26         & 0.23       & 0.20       & 0.16        & 0.13 \\
    \cline{3-11}
    0.1     & 30 & 0.37        & 0.31       & 0.30        & 0.28         & 0.24         & 0.21       & 0.18       & 0.15        & 0.11 \\
            &  0 & 0.37        & 0.32       & 0.29        & 0.27         & 0.24         & 0.21       & 0.18       & 0.15        & 0.11 \\
    \cline{3-11}
    0.2     & 30 & 0.32        & 0.27       & 0.26        & 0.23         & 0.20         & 0.19       & 0.16       & 0.13        & 0.10  \\
            &  0 & 0.32        & 0.27       & 0.25        & 0.23         & 0.21         & 0.19       & 0.16       & 0.13        & 0.10  \\
    \cline{3-11}
    0.3     & 30 &  0.26       & 0.24       & 0.21        & 0.19         & 0.18         & 0.16       & 0.14       & 0.12        & 0.09 \\
            &  0 & 0.28        & 0.24       & 0.21        & 0.19         & 0.18         & 0.16       & 0.14       & 0.12        & 0.09 \\
    \cline{3-11}
    0.4     & 30 & 0.21        & 0.20       & 0.18        & 0.16         & 0.14         & 0.13       & 0.12       & 0.10       & 0.07 \\
            &  0 & 0.23        & 0.20       & 0.18        & 0.16         & 0.14         & 0.13       & 0.11       & 0.10       & 0.07 \\
    \cline{3-11}
    0.5     & 30 & 0.16        & 0.15       & 0.14        & 0.13         & 0.12         & 0.10       & 0.09        & 0.08       & 0.06  \\
            & 0  & 0.18        & 0.16       & 0.14        & 0.13         & 0.12         & 0.10       & 0.09        & 0.08       & 0.06  \\
    \cline{3-11}
    0.6     & 30 & 0.12        & 0.11       & 0.10        & 0.09         & 0.09         & 0.08       & 0.07        & 0.06       & 0.050   \\
            &  0 & 0.13        & 0.12       & 0.11        & 0.10         & 0.09         & 0.08       & 0.07        & 0.06       & 0.045 \\
    \cline{3-11}
    0.7     & 30 & 0.08        & 0.08       & 0.07        & 0.07         & 0.06         & 0.05       & 0.05        & 0.045      & 0.035 \\
            &  0 & 0.09        & 0.08       & 0.07        & 0.06         & 0.06         & 0.05       & 0.05        & 0.045      & 0.035 \\
    \cline{3-11}
    0.8     & 30 & 0.05        & 0.04       & 0.04        & 0.04         & 0.04         & 0.035      & 0.03        & 0.025      & 0.0235  \\
            &  0 & 0.05        & 0.05       & 0.04        & 0.04         & 0.04         & 0.035      & 0.03        & 0.025      & 0.0230 \\
    \hline
    \end{tabular}
    \caption{Critical semi-major axis $a_c$  for a planet  mass ($m$) of 30 Jupiter mass (M$_\textrm{J}$), and for a planet of negligible mass (test particle), as a function of the mass ratio $\mu$ and of the eccentricity $e$ of the binary.}
    \label{tab:1}
\end{table*}

A relevant parameter is the so-called \textit{critical} semi-major axis $a_c$, defined as the edge that separates the stable and unstable region; this idea assumes that all the orbits with $a \leq a_c$ are stable and all orbits with $a > a_c$ are unstable.  We notice that $a_c$ is well defined only if the boundary between the region of stability (where $a \le a_c$) and instability where ($a > a_c$) is sharp. 

Table \ref{tab:1} reports $a_c$ as a function of the mass ratio $\mu$ (from 0.1 to 0.9) and eccentricity $e$ (from 0 to 0.8) of the binary star system for test particles - i.e. planets with negligible mass - and for 30 Jupiter mass planets.

 The results presented assume that the planetary system is isolated. However, for binary system systems living in star clusters, the effect of scattering with a passing by star must in general be taken into account. To attack this problem, we performed another large number (more than 4500) of scattering experiments with different initial conditions typical of encounters in small star clusters. Figure \ref{fig:6} shows an example of a trajectory in a system that has been perturbed by the scattering of a passing by star. In this case, the trajectory shows that the planet is temporarily caught by the secondary star (whose orbit is in red).

\section{Conclusions}
    The problem of stability in binary star systems is a longstanding topic. In our work we have made several progresses compared to previous works. In particular:

(i) we apply a high precision numerical integrator, also considering planets of not negligible mass; (ii) we evaluate the statistics of the various types of instability, the ejections of the planets and the collisions with the primary or secondary star; (iii) we study the effect of the external perturbations due to a passing by star in a star cluster.\\
In the future, we will go in different directions, taking into account that the application of High Performance Computing (HPC) techniques will allow a significant extension of the space of parameters. In particular, we will investigate the dependence of stability on the inclination of the orbital plane of the planet and on its initial eccentricity. Data analysis may be challenging in the case of very large dataset; a promising way to get results for data is based on Machine Learning techniques (see \cite{ta16}).
    
\bibliographystyle{mnras}
\bibliography{ref}
\end{document}